\documentstyle[aps,prd,eqsecnum,floats]{revtex}
\input{psfig}

\begin{document}
\draft
\def \tablerule {\noalign{\hrule}}


\title{
Matched filtering of gravitational waves from inspiraling compact
binaries: Computational cost and template placement}
\author{Benjamin J. Owen$^1$ and B.~S. Sathyaprakash$^{1,2}$}
\address{
$^1$Theoretical Astrophysics, California Institute of Technology,
Pasadena, California 91125}
\address{
$^2$Department of Physics and Astronomy, Cardiff University, Cardiff, 
CF2 3YB, United Kingdom}
\date{August 28, 1998}
\maketitle

\begin{abstract}
We estimate the number of templates, computational power, and storage
required for a one-step matched filtering search for gravitational
waves from inspiraling compact binaries.  Our estimates for the
one-step search strategy should serve as benchmarks for the evaluation
of more sophisticated strategies such as hierarchical searches.  We
use a discrete family of two-parameter waveform templates based on the
second post-Newtonian approximation for binaries composed of
nonspinning compact bodies in circular orbits.  We present estimates
for all of the large- and mid-scale interferometers now under
construction: LIGO (three configurations), VIRGO, GEO600, and TAMA.
To search for binaries with components more massive than $m_{\min} =
0.2M_\odot$ while losing no more than $10\%$ of events due to
coarseness of template spacing, the initial LIGO interferometers will
require about $1.0 \times 10^{11}$ flops (floating point operations
per second) for data analysis to keep up with data acquisition.  This
is several times higher than estimated in previous work by Owen, in
part because of the improved family of templates and in part because
we use more realistic (higher) sampling rates.  Enhanced LIGO, GEO600,
and TAMA will require computational power similar to initial LIGO.
Advanced LIGO will require $7.8 \times 10^{11}$ flops, and VIRGO will
require $4.8 \times 10^{12}$ flops to take full advantage of its broad
target noise spectrum.  If the templates are stored rather than
generated as needed, storage requirements range from $1.5 \times
10^{11}$ real numbers for TAMA to $6.2 \times 10^{14}$ for VIRGO.  The
computational power required scales roughly as $m_{\min}^{-8/3}$ and
the storage as $m_{\min}^{-13/3}$.  Since these scalings are perturbed
by the curvature of the parameter space at second post-Newtonian
order, we also provide estimates for a search with $m_{\min} =
1M_\odot$.  Finally, we sketch and discuss an algorithm for placing
the templates in the parameter space.
\end{abstract}


\section{Introduction} 
\label {sec:introduction}

Close binary systems composed of compact objects (such as black holes
and neutron stars) are expected to be an important source of
gravitational waves for broadband laser interferometers such as LIGO,
VIRGO, GEO600, and TAMA~\cite{LIGO,VIRGO,GEO,TAMA}.  The orbit of a
compact binary decays under the influence of gravitational radiation
reaction, emitting a gravitational wave signal that increases in
amplitude and ``chirps'' upward in frequency as the objects spiral in
toward each other just before their final coalescence.  According to
current astronomical lore~\cite{lore,optimists,pessimists}, the rate
of coalescences should be about three per year within 200 to
300~Mpc~\cite{lore} of the Earth for neutron-star--neutron-star
binaries, and within 400~Mpc to 1~Gpc for black-hole--black-hole
binaries.  Signals from inspiraling compact binaries at these
distances are strong enough to be detected by ``enhanced''
LIGO~\cite{enhanced} and VIRGO interferometers, but only if aided by a
nearly optimal signal-processing technique.  Fortunately, the
distinctive frequency chirp has been calculated to a remarkable degree
of precision using a variety of approximations to the general
relativistic two-body problem (e.g.~\cite{BDIWW,TSTS96}).  Because the
functional form of the chirp is quite well-known, a search for
inspiral signals in noisy data is ideally suited to matched filtering.

Matched filtering~\cite{Wiener} has long been known to be the optimal
linear signal-processing technique and is well-discussed in the
literature (e.g.~\cite{Helstrom}); therefore we will only briefly
summarize it here.  In the frequency domain, a matched filter is a
best-guess template of the expected signal waveform divided by the
interferometer's spectral noise density.  The interferometer output is
cross-correlated with the matched filter at different time delays to
produce a filtered output.  The signal-to-noise ratio, defined as the
ratio of the actual value of the filtered output to its rms value in
the presence of pure noise, is compared to a predetermined threshold
to decide if a signal is present in the noise.  If the signal from
which the matched filter was constructed is present, it contributes
coherently to the cross-correlation, while the noise contributes
incoherently and thus is reduced relative to the signal.  Also, the
weighting of the cross-correlation by the inverse of the spectral
noise density emphasizes those frequencies to which the interferometer
is most sensitive.  Consequently, signals thousands of cycles long
whose unfiltered amplitude is only a few percent of the rms noise can
be detected.

A matched filtering search for inspiraling compact binaries can be
computationally intensive due to the variety of possible waveforms.
Although the inspiral signals are all expected to have the same
functional form, this form depends on several parameters---the masses
of the two objects, their spins, the eccentricity of their orbit,
etc.---some of them quite strongly.  A filter constructed from a
waveform template with any given parameter vector may do a very poor
job of detecting a signal with another parameter vector.  That is, the
difference in parameter vectors can lead to a greatly reduced
cross-correlation between the two wave forms; and in general, the
greater the difference, the more the cross-correlation is reduced.
Because the parameter vector of a signal is not known in advance, it
is necessary to filter the data with a family of templates located at
various points in parameter space---e.g., placed on a lattice---such
that any signal will lie close enough to at least one of the templates
to have a good cross-correlation with that template.

There are several questions that must be answered in order to
determine the feasibility of a matched filtering search strategy and,
if feasible, to implement it.  Which parameters significantly affect
the waveform?  How should the spacing of the template parameters
(lattice points) be chosen?  Is there a parametrization that is in
some sense ``preferred'' by the template waveforms?  How many
templates are needed to cover a given region of interest in the
parameter space, and how much computing power and memory will it cost
to process the data through them?  In the case of inspiraling compact
binaries, the full general-relativistic waveforms are not exactly
known, but are instead approximated (e.g., using the post-Newtonian
scheme); and we must also ask, what approximation to the true
waveforms is good enough?

All of these questions have been addressed in recent years, at least
at some level.  The current state of affairs is summarized by the following
brief review of the literature:  

The standard measure for deciding what class of waveforms is good
enough is the {\it fitting factor} ($FF$) introduced by
Apostolatos~\cite{A95}.  The fitting factor is effectively the
fraction of optimal signal-to-noise-ratio obtained when filtering the
data with an approximate family of templates.  Because binaries are
(on large scales) uniformly distributed in space and because the
signal strength scales inversely with distance, the fraction of event
rate retained is approximately $FF^3$.  Therefore it has become
conventional to regard $FF = 97\%$---i.e., $10\%$ loss of event
rate---as a reasonable goal.  Using the standard post-Newtonian
expansion in the test-mass case (i.e., when one body is much less
massive than the other so that the waveforms can be computed with
arbitrarily high precision using the Teukolsky formalism), Droz and
Poisson~\cite{DP97} found that 2PN signals had fitting factors of
$90\%$ or higher.  Damour, Iyer, and Sathyaprakash~\cite{DIS98} have
devised a new way of rearranging the usual post-Newtonian expansion
(similar to the way Pad\'e approximants rearrange the coefficients of
a Taylor expansion) to take advantage of physical intuition in
constructing templates.  They find fitting factors of $95\%$ or higher
for the 2PN templates.  Research underway by the authors
of~\cite{BDIWW} will lead to 3PN templates that should easily achieve
$FF>97\%$.

Several people~\cite{BD94,A95,KKS95} have shown that it is
insufficient to use templates that depend on just one shape parameter
(the ``chirp mass,'' which governs the rate of frequency sweep at
Newtonian-quadrupole order).  To achieve $FF > 90\%$ one must include
the masses of both objects as template parameters, as was done in the
above 2PN analyses~\cite{DP97,DIS98}, and as is being done in the
forthcoming 3PN templates.

Apostolatos~\cite{A95,A96} showed that, for binaries whose components
spin rapidly about their own axes which are orthogonal to the orbital
plane so that there is no precession, neglecting the spin parameters (i.e.,
using two-mass-parameter waveforms based on the theory of spinless
binaries) degraded the fitting factors by less than $2\%$.  With
precession the situation is much more complicated, and data analysis
algorithms are as yet poorly developed: It is clear that there are
interesting corners of parameter space (most especially a neutron star
in a substantially nonequatorial, precessing orbit around a much more
massive, rapidly spinning black hole) in which the two-mass-parameter
spinless waveforms give $FF \ll 90\%$; to search for such binaries
will require waveforms with three or more parameters~\cite{A96}.
However, the 2PN (or 3PN) two-mass-parameter waveforms do appear to
cover adequately a significant portion of the parameter space for
precessing binaries~\cite{A95}.

Sathyaprakash~\cite{Sathya94} showed that in computations with the
two-mass-parameter waveforms, the best coordinates to use on the
parameter space are not the two masses, but rather the inspiral times
from some fiducial frequency to final merger, as computed at Newtonian
and first post-Newtonian order. Working with the {\it restricted}
first post-Newtonian wave forms (see below) he found that the
effective dimension of the parameter space is nearly one.

Sathyaprakash and Dhurandhar~\cite{SD91,SD93,DS94} developed a
criterion for putting templates at discrete points on a grid in
parameter space and numerically implemented their criterion for a
one-parameter (Newtonian) family of templates and for simple noise
models.  They introduced the concept of what Owen~\cite{Owen96} later
called the {\it minimal match} (analagous to the fitting factor) as a
measure of how well a discrete set of templates covers the parameter
space and estimated the computational costs for an on-line search.

Owen~\cite{Owen96}, building on the work of Sathyaprakash and
Dhurandhar, defined a metric on the parameter space from which one can
semi-analytically calculate (i) the template spacing needed to achieve
a desired minimal match, (ii) the total number of templates needed,
and (iii) the computational requirements to keep up with the
data---for any family of waveforms and any interferometer noise
spectrum.  Owen combined this metric-based formalism with
computational counting procedures from Schutz~\cite{Schutz91} to
estimate the computational requirements for LIGO searches based on
two-parameter 1PN templates.  These estimates were confirmed by
Apostolatos~\cite{A96} using a numerical method in the vein of (but
more sophisticated than) the previous work of Sathyaprakash and
Durandhar~\cite{SD91,SD93,DS94}.  Apostolatos also showed that a
search for precessing binaries that fully covers all the nooks and
crannies of the precessional parameter space, using currently
available templates and techniques, is prohibitively costly.

Mohanty and Dhurandhar~\cite{MD96,Mohanty98} have studied hierarchical
search strategies.  Such strategies reduce computational costs by
making a first pass of the data through a coarsely-spaced template
grid and a low signal-to-noise threshold to identify candidate
signals.  Each candidate flagged by the first pass is examined more
closely with a second, finely-spaced grid of templates and a higher
threshold to weed out false alarms.  Such strategies can reduce the
total computational requirements by roughly a factor 25.

The purpose of this paper is to refine and update the analyses by
Owen~\cite{Owen96} for the two-parameter, spinless templates that are
likely to be used for binary-inspiral searches in ground-based
interferometers.  This refinement is needed because the
kilometer-scale interferometers will begin taking data in about 2
years (preliminary, engineering run); people are even now designing
software to implement the simplest matched filtering search algorithm;
and in the context of these implementations, the factor of 3 accuracy
attempted in Ref.~\cite{Owen96} is no longer adequate.  The numbers
that are derived in this paper should establish a reliable baseline
cost to which more sophisticated search strategies (e.g., hierarchical
searches) can be compared.

The substantial differences between this paper and Ref.~\cite{Owen96}
are that we now (i) approximate the phase evolution of the inspiral
waveform to 2PN rather than 1PN order; (ii) give results for the noise
spectra of several more interferometers; and (iii) use a better
estimate of the sampling frequency needed for each interferometer.  We
assume the following fiducial search: a minimal match of $97\%$
(corresponding to $10\%$ loss of event rate due to coarse parameter
space coverage), second post-Newtonian waveforms, and templates made
for objects of minimum mass $m_{\min} = 0.2M_\odot$ and up.

Our results for the computational requirements are given in
Tables~\ref{table:calN}--\ref{table:cost2}.  These tables show that
the initial LIGO interferometers need about twice as many templates
and triple the computational power estimated in Ref.~\cite{Owen96}.
These increases result mainly from using 2PN waveforms rather than the
(clearly inadequate) 1PN, and from a using a higher sampling rate (as,
it turns out, is required to keep time-step discretization error from
compromising the $97\%$ minimal match).  GEO600 requires slightly more
templates and power than LIGO because of its flatter noise spectrum,
while TAMA requires slightly less because its sensitivity is limited
to higher frequencies where there are fewer cycles.  Initial VIRGO,
with its extremely broad and flat spectrum, requires about the same as
advanced LIGO.  We conclude that even if VIRGO achieves its targeted
low-frequency sensitivity, it will be unable to take full advantage of
that sensitivity with matched filtering because of the large number of
templates required.

The rest of this paper is organized as follows.  In Sec.~II we analyze
the application of matched filtering to a search for inspiraling
binaries and summarize the method of Ref.~\cite{Owen96} which uses
differential geometry to answer important questions about such a
search.  We use this method in Sec.~III to estimate the computational
costs and other requirements of a matched-filtering binary search for
LIGO, VIRGO, GEO600, and TAMA.  In Sec.~IV we illustrate a detailed
example of a template placement algorithm, and in Sec.~V we discuss
our results.

\section{Formalism} 
\label {sec:formalism}

This Section summarizes material previously presented in~\cite{Owen96}
with several incremental improvements.  We begin by introducing some
notation.

The Fourier transform of a function $h(t)$ is denoted by
$\tilde{h}(f)$, where
\begin{equation}
\tilde{h}(f)\equiv\int_{-\infty}^\infty dt\,e^{i2\pi ft}h(t).
\end{equation}
We write the interferometer output $h(t)$ as the sum of noise $n(t)$
and a signal ${\cal A}s(t)$, where we have separated the signal into a
dimensionless, time-independent amplitude ${\cal A}$ and a ``shape''
function $s(t)$ which is defined to have unit norm [see
Eq.~(\ref{def:ip}) below].

The strain power spectral noise density of an interferometer is
denoted by $S_h(f)$.  We use the one-sided spectral density, defined
by
\begin{equation}
\mbox{E}[\tilde{n}(f_1)\tilde{n}^*(f_2)]
=\frac{1}{2}\delta(f_1-f_2)S_h(|f_1|),
\end{equation}
where $\mbox{E}[~]$ denotes the expectation value over an ensemble of
realizations of the noise and an asterisk denotes
complex conjugation.

We use geometrized units, i.e.\ Newton's gravitational constant $G$
and the speed of light $c$ have values of unity.

\subsection{Matched filtering} 
\label {sec:matched filtering}

First we flesh out the Introduction's brief description of matched
filtering.  In the simplest idealization of matched filtering, the
filtered signal-to-noise ratio is defined by \cite{Helstrom}
\begin{equation}
\label{def:SNR}
{S\over N}\equiv\frac{\left<h,u\right>}{\mbox{rms }\left<n,u\right>}.
\end{equation}
Here $u$ is the template waveform used to filter the data stream $h$,
and the inner product
\begin{equation}
\label{def:ip}
\left<a,b\right>\equiv
4\,\mbox{Re}\left[\int_0^\infty df\,\frac{\tilde{a}^*(f)\,\tilde{b}(f)}
{S_h(f)}\right]
\end{equation}
is the noise-weighted cross-correlation between $a$ and $b$
(cf.~\cite{CF94}).  The denominator of~(\ref{def:SNR}) is equal to
$\sqrt{\left<u,u\right>}$, the norm of $u$ (see Sec.~II~B of
Ref.~\cite{CF94} for a proof).  Because the norm of $u$ cancels out
of~(\ref{def:SNR}), we can simplify our calculations without loss of
generality by considering all templates to have unit norm.

When searching for a parametrized family of signals the situation is
somewhat more complicated.  The parameter values of the signals are
not known in advance; therefore one must filter the data through many
templates constructed at different points in the parameter space.  To
develop a strategy for searching the parameter space, one must know
how much the $S/N$ is reduced by using a template whose parameter
values differ from those of the signal.  Neglecting fluctuations due
to the noise, the fraction of the optimal $S/N$ obtained by using the
wrong parameter values is given by the {\it ambiguity function},
\begin{equation}
\label{def:A}
A(\bbox{\lambda},\bbox{\Lambda}) \equiv
\left< u(\bbox{\lambda}), \,u(\bbox{\Lambda}) \right>
\end{equation}
(see, e.g., Chs.~XIII and~X of Ref.~\cite{Helstrom}).  Here
$\bbox{\lambda}$ and $\bbox{\Lambda}$ are the parameter vectors of the
signal and template (it doesn't matter which is which).  The ambiguity
function, as its name implies, is a measure of how distinguishable two
waveforms are with respect to the matched filtering process.  It can
be regarded as an inner product on the waveform parameter space and is
fundamental to the theory of parameter estimation~\cite{Helstrom,BSD96}.

For the purposes of a search for inspiraling compact binaries, the
ambiguity function isn't quite what is needed.  This is because the
test statistic (for a given set of parameter values $\bbox{\theta}$)
is not given by Eq.~(\ref{def:SNR}), but rather by
\begin{equation}
\label{def:statistic}
\renewcommand{\arraystretch}{.6}
\begin{array}[t]{c}{\textstyle\max}\\{\scriptstyle\phi_c,t_c}\end{array}
{\langle h,u(\bbox{\theta}) e^{i(2\pi ft_c-\phi_c)}\rangle
\over\mbox{rms }\langle n,u(\bbox{\theta})\rangle}.
\end{equation}
Here $\phi_c$ and $t_c$ are respectively the coalescence time and
coalescence phase.  We separate these parameters out from the rest:
$\bbox{\lambda}=(\phi_c,t_c,\bbox{\theta})$, where $\bbox{\theta}$ is
the vector of {\it intrinsic parameters} that determine the shape of
the waveform and $\phi_c$ and $t_c$ are {\it extrinsic
parameters}~\cite{Owen96} (also refereed to as {\it kinematical}
and {\it dynamical} parameters \cite{Sathya94}, respectively).  
The practical difference is that
maximization over the extrinsic parameters is performed automatically
by Fourier transforming, taking the absolute value, and looking for
peaks.  The use of Eq.~(\ref{def:statistic}) as a detection statistic
suggests the definition of a modified ambiguity function known as the
{\it match}~\cite{Owen96}
\begin{equation}
\label{def:match}
M(\bbox{\theta}_1,\bbox{\theta}_2)\equiv
\renewcommand{\arraystretch}{.6}
\begin{array}[t]{c}{\textstyle\max}\\{\scriptstyle\phi_c,t_c}\end{array}
\langle u(\bbox{\theta}_1),u(\bbox{\theta}_2) e^{i(2\pi ft_c-\phi_c)}
\rangle,
\end{equation}
where the templates $u$ are assumed to have unit norm.  The use of
this match function rather than the ambiguity function takes into
account the fact that a search can benefit from systematic errors in
the extrinsic parameters.

\subsection{Applications of differential geometry} 
\label {sec:differential geometry}

The match (\ref{def:match}) can be regarded as an inner product on the
space of template shapes and intrinsic template parameters, and
correspondingly one can define a metric on this space~\cite{Owen96}:
\begin{equation}
\label{def:metric}
g_{ij}(\bbox{\theta})\equiv
-\frac{1}{2}\left.\frac{\partial^2M(\bbox{\theta},\bbox{\Theta})}
{\partial\Theta^i\partial\Theta^j}\right|_{\Theta^k=\theta^k}.
\end{equation}
The metric~(\ref{def:metric}) is derived from the
match~(\ref{def:match}) in the same way the information matrix
$\Gamma_{ij}$ is derived from the ambiguity function~\cite{BSD96}, and
plays a role in signal detection similar to that played by the
information matrix in parameter estimation.  The $g_{ij}$ can be
derived by expanding $M(\bbox{\theta},\bbox{\Theta})$ about
$\bbox{\Theta}=\bbox{\theta}$, or equivalently by projecting
$\Gamma_{ij}$ on the subspace orthogonal to $\phi_c$ and $t_c$.

\begin{figure}
\centerline{\psfig{file=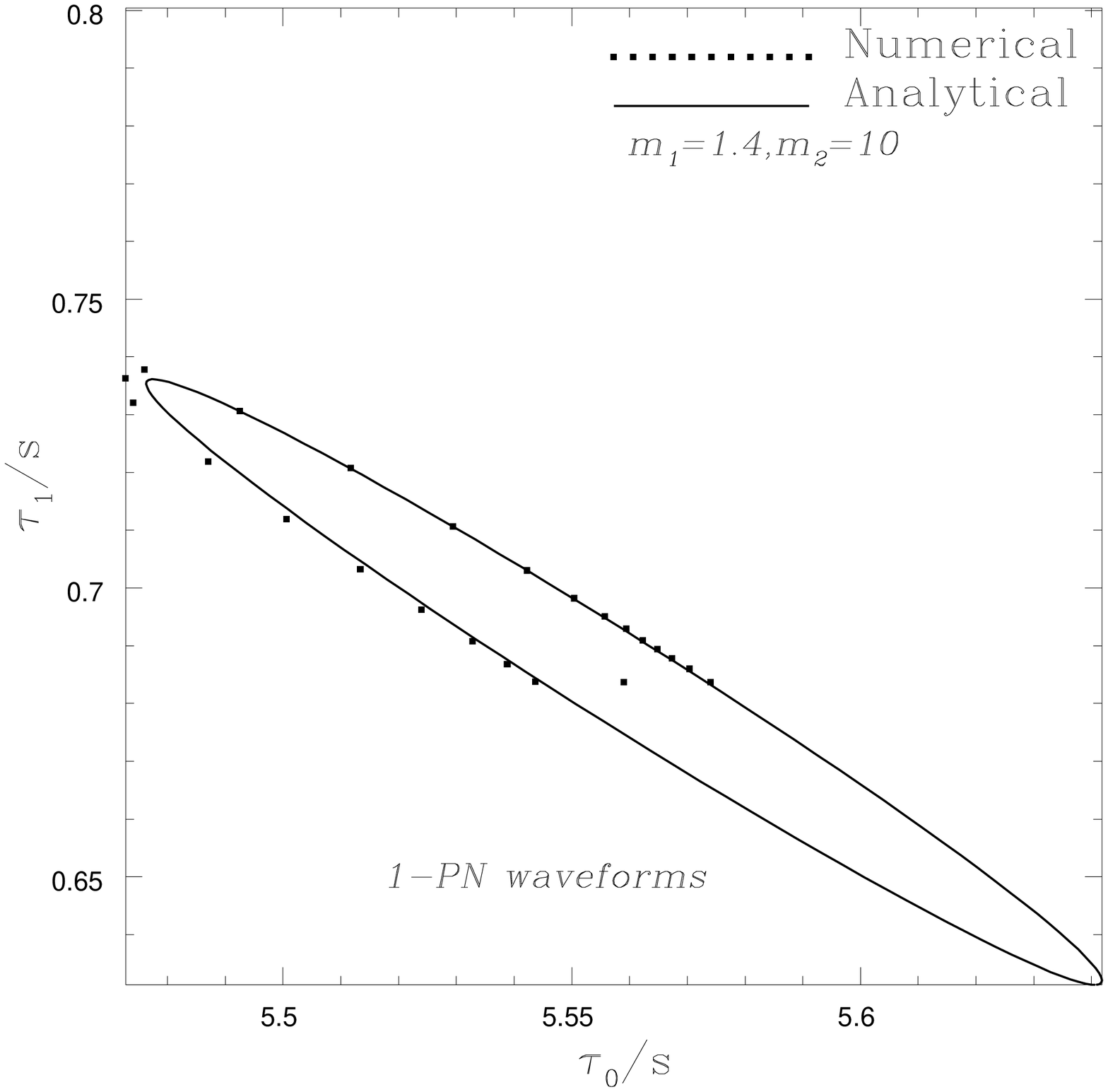,height=3.in}}
\centerline{\psfig{file=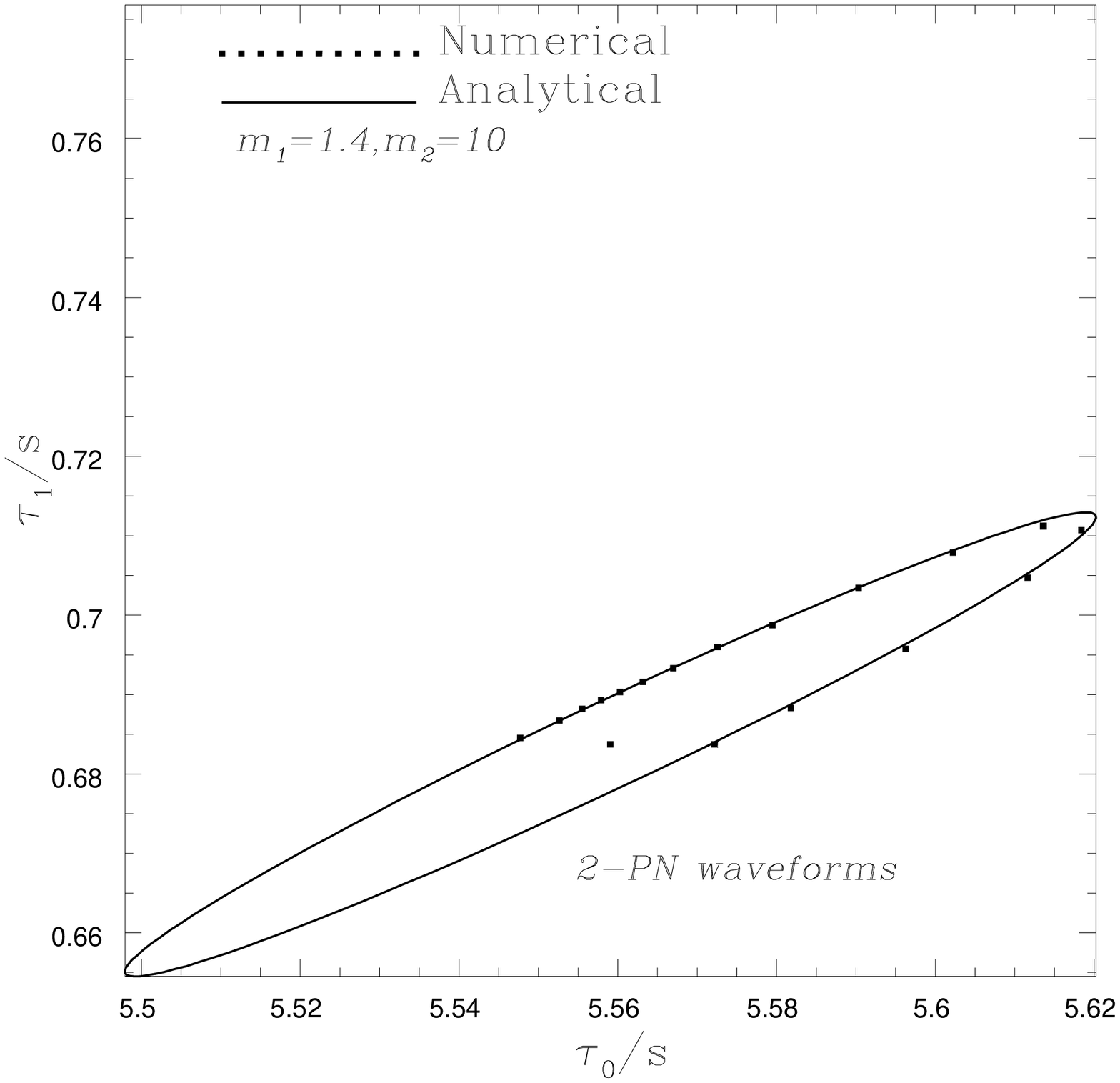,height=3.in}}
\caption{
Comparison of the match to the quadratic approximation in the case of:
(a) first post-Newtonian waveforms and (b) second post-Newtonian wave
forms.  The solid line is obtained using the analytical model
Eq.~(\protect\ref{match approx}) and the dots denote the match
obtained by numerically computing the `distance' from the centre, at
which the match drops to the chosen value 0.97.  The two agree very
well in the case of first post-Newtonian waveforms while in the second
post-Newtonian case the agreement is quite good for all but waveforms
of massive binaries.  The coordinates $(\tau_0, \tau_1)$ are defined
in Eq.~(\protect{\ref{taudefs}}).}
\label{fig:quadratic}
\end{figure}

The $g_{ij}$ can be used to approximate the match in the regime
$1-M\ll 1$ by
\begin{equation}
\label{match approx}
M(\bbox{\theta},\bbox{\theta}+\Delta\bbox{\theta})
\simeq 1-g_{ij}\Delta\theta^i\Delta\theta^j,
\end{equation}
which is simply another way of writing the Taylor expansion of
$M(\bbox{\theta},\bbox{\theta}+\Delta\bbox{\theta})$ about
$\Delta\bbox{\theta}=0$.  (The first derivative term vanishes because
$M$ takes its maximum value of unity at $\Delta\bbox{\theta}=0$.)  We
find that the quadratic approximation~(\ref{match approx}) is good
typically for $M\simeq 0.95$ or greater, though this depends on the
waveform and noise spectrum used.  Experience suggests that the
quadratic approximation generally underestimates the true match; and
thus the spacings and numbers of templates we calculate using
Eq.~(\ref{match approx}) err on the safe side.  See
Fig.~\ref{fig:quadratic} for an example.

In the limit of close template spacing, Eq.~(\ref{match approx}) leads
to a simple, analytical way of placing templates on a lattice.  We
discuss this in some detail in Sec.~\ref{sec:placement}, but for now
turn our attention to the use of the quadratic approximation in
calculating the number of templates needed for a lattice.

\subsection{Computational costs}
\label {sec:number of templates theory}

If the number ${\cal N}$ of templates needed to cover a region of
interest is large, it is well approximated by the ratio of the proper
volume of the region of interest to the proper volume per template
$\Delta V$,
\begin{equation}
\label{eq:calN}
{\cal N}=(\Delta V)^{-1} \int d^D\theta \,\sqrt{\det \|g_{ij}\|},
\end{equation}
where $D$ is the dimension of the parameter space~\cite{Owen96}.
Equation~(\ref{eq:calN}) underestimates ${\cal N}$ when not in the
limit $\Delta V \to 0$ ( ${\cal N} \to \infty$).  The reason is
template {\it spill over,} i.e.\ the fact that in any real algorithm
for laying out templates, those on the boundaries of the region of
interest will to some extent cover regions just outside.  This effect
is small in the limit of many templates because it goes as the
surface-to-volume ratio of the region of interest.

The proper volume per template, $\Delta V$, depends on the packing
algorithm used, which in turn depends on the number $D$ of dimensions
(see Sec.~\ref{sec:placement}).  For $D=2$, the optimal packing is a
hexagonal lattice, and thus
\begin{equation}
\label{eq:hex}
\Delta V=\frac{3\sqrt{3}}{2}(1-MM),
\end{equation}
where $MM$ is the {\it minimal match} parameter defined in
Ref.~\cite{Owen96} as the match between signal and template in the
case when the signal lies equidistant between all the nearest
templates (i.e., the worst-case scenario).  There is no packing scheme
which is optimal for all $D$, but it is always possible (though
inefficient) to use a hypercubic lattice, for which
\begin{equation}
\label{eq:dV}
\Delta V=(2\sqrt{(1-MM)/D})^D.
\end{equation}

For inspiraling compact binaries, Ref.~\cite{Owen96} has spelled out a
detailed prescription for obtaining the $g_{ij}$ needed to evaluate
the proper volume integral in Eq.~(\ref{eq:calN}).  In summary, first
one obtains a metric including the $t_c$ parameter,
\begin{equation}
\label{gamma_ab}
\gamma_{\alpha\beta}=\frac{1}{2}({\cal J}[\psi_\alpha\psi_\beta]
-{\cal J}[\psi_\alpha]{\cal J}[\psi_\beta]),
\end{equation}
where $\psi_\alpha$ is the gradient of the waveform phase $\Psi$ in
the parameter space of intrinsic parameters plus $t_c$ and the moment
functionals
\begin{equation}
\label{calJa}
{\cal J}[a]\equiv \frac {\langle f^{-7/3},a(f)\rangle}
{4\int_0^\infty df\,{f^{-7/3}\over S_h(f)}}
\end{equation}
can be expanded (for binary chirp waveforms) in terms of the noise
moments~\cite{PW95}
\begin{equation}
\label{Jp}
J(p)= \left [\int_0^\infty df\, \frac{(f/f_0)^{-7/3}}{S_h(f)}
\right]^{-1} \int_0^\infty df\, \frac{(f/f_0)^{-p/3}}{S_h(f)}
\end{equation}
where $f_0$ is the frequency of the minimum of $S_h(f)$.  Then one
projects out the coalescence time $t_c$ to obtain
\begin{equation}
\label{project}
g_{ij}=\gamma_{ij}-\gamma_{0i}\gamma_{0j}/\gamma_{00}.
\end{equation}

Once ${\cal N}$ has been found it is a relatively straightforward
matter to calculate the CPU power and storage required to process all
the templates in an on-line search.  The interferometer data stream is
broken up for processing into segments of $D$ samples (real numbers),
such that $D \gg F$ where $F$ is the length (in real numbers) of the
longest filter.  (See Schutz~\cite{Schutz91} for a discussion of the
optimization of $D/F$, taking into account the fact that successive
data segments must overlap by at least $F$ to avoid circular
correlations in the Fourier transform.)  Using the operations count
for a real Fourier transform~\cite{Schutz91}, filtering the data
segment through ${\cal N}$ templates of length $F$ requires
\begin{equation}
{\cal N}\,D(16+3\log_2F)
\end{equation}
floating point operations.  If we take the sampling frequency
to be $2f_u$ (see Sec.~III and Table~\ref{table:noise}), the
computational power required to keep pace with data acquisition is
\begin{equation}
\label{eq:calP}
{\cal P} \simeq {\cal N}\,f_u (32 + 6\log_2 F)
\end{equation}
flops (floating point operations per second).

\section{Computational cost using restricted 2PN templates}
\label {sec:computational cost 2pn}

In this section, using the geometric formalism summarized in Sec.~II,
we calculate the number ${\cal N}$ of templates required to cover a
region of interest as a function of the minimal match.  We then use
this number to calculate the computational cost of filtering a single
interferometer's output through all these templates in an on-line
search.

\subsection{Functional form of the templates}

We construct our waveform templates using two intrinsic parameters
based on the masses of the binary's components.  Inspiral waveforms in
principle can be strongly affected by several other parameters: spins
of the two components, orbital eccentricity, and several angles
describing the orientation of the binary with respect to the
interferometer.  However, it is believed that two-parameter templates
will be adequate to search for most binaries for the following
reasons.

(i) Based on models of the evolution of currently known binary
pulsars, it is expected~\cite{BDIWW} that typical NS-NS binaries will
have spins of negligible magnitude (spin/mass$^2 \ll 1$).
Apostolatos~\cite{A96} has shown that, even if the magnitudes of the
spins are large, their effect on $S/N$ is small (reduces $FF$ by less
than $2\%$ if the orbit and the spin vectors do not precess).  He has
also shown that precession will not reduce $FF$ below $90\%$ except in
the relatively small region of parameter space containing binaries
with a neutron star orbiting a more massive, rapidly rotating
(spin/mass$^2 \sim 1$) black hole with orbital angular momentum
inclined by more than about 30 degrees to the black hole's spin.

(ii) It has long been known~\cite{eccentricity} that gravitational
radiation reaction circularizes all but the most eccentric orbits on a
time scale much smaller than the lifetime of the binary if the
progenitor system was the same binary.  This may not be true, however,
in the case of close binaries formed by capture in densely populated
environments, e.g.\ galactic nuclei/globular clusters.

(iii) The angles make no difference in our analysis because we use the
{\it restricted post-Newtonian} approximation~\cite{CF94}, in which
the phase evolution of the inspiral waveform is followed to a high
post-Newtonian order but the amplitude is only followed to lowest
order.  In this approximation, the combined effect of the angles is to
multiply the waveform by a constant amplitude and phase factor, which
does not affect the choice of search templates \cite{SD91}.  Presently it is
believed~\cite{CF94} that the restricted post-Newtonian approximation
will be good enough for data analysis of ground-based interferometers.

The standard post-Newtonian expansion of the waveform phase is given
as a function of mass parameters based on the standard astronomical
choices $M$ (total mass) and $\mu$ (reduced mass).  In order to
clearly isolate test-mass terms (i.e., those that remain when one body
is much less massive than the other), the symmetric mass ratio $\eta =
\mu/M$ is typically used instead of $\mu$.  In terms of $M$ and
$\eta$, the second post-Newtonian waveform phase can be calculated
from the energy loss formula of Blanchet {\it et al.}~\cite{BDIWW} as
\begin{eqnarray}
\label{Psi}
\Psi(f;M,\eta) &=& {3\over128} (\pi Mf)^{-5/3} \left[1 +{20\over9}
\left( {743\over336} +{11\over4} \eta \right) (\pi Mf)^{2/3}
-16\pi(\pi Mf) \right.\nonumber\\
&&\left. +10\left( {3\,058\,673 \over 1\,016\,064}
+{5\,429 \over 1\,008} \eta +{617\over144} \eta^2
\right) (\pi Mf)^{4/3} \right]
\end{eqnarray}
[cf.\ Eq.~(3.6) of Poisson and Will~\cite{PW95}].  However, $M$ and
$\eta$ are inconvenient parameters for our purposes because, when they
are used as the parameter-space coordinates, the values of the metric
components vary strongly over parameter space, making calculations
unnecessarily difficult and prone to numerical error.

In earlier analyses~\cite{Sathya94,BSD96,Owen96} it was found more
convenient to use as parameters the Newtonian and 1PN {\it chirp
times}
\begin{mathletters}
\label{taudefs}
\begin{eqnarray}
\tau_0 &=& {5\over256} M^{-5/3} (\pi f_0)^{-8/3} \eta^{-1},\\
\tau_1 &=& {5\over192} M^{-1} (\pi f_0)^{-2} \left({743\over336\eta}
+{11\over4}\right),
\end{eqnarray}
\end{mathletters}
which are respectively the Newtonian and 1PN contributions to the time
it takes the instantaneous gravitational-wave frequency to (formally)
evolve from $f_0$ to infinity.  The chirp times are more convenient
than the usual mass parameters because, when they are chosen as
parameter-space coordinates, at 1PN order the metric components are
constant.  Assuming the post-Newtonian expansion has reasonable
convergence properties, one would expect the metric components in
these coordinates to remain nearly constant at higher post-Newtonian
orders (and indeed we find this is so).  However, at higher than 1PN
order one cannot write the waveform phase analytically in terms of
$\tau_0$ and $\tau_1$.  To remedy this, following
Mohanty~\cite{Mohanty98}, we base our second parameter on the 1.5PN
chirp time $\tau_{1.5}$ (see \cite{BSD96} for a definition).  More 
specifically, we introduce new
dimensionless coordinates in parameter space.  We define
\begin{mathletters}
\begin{eqnarray}
\theta^1 &=& 2\pi f_0\tau_0 = {5\over128}(\pi Mf_0)^{-5/3}\eta^{-1},\\
\theta^2 &=& 2\pi f_0\tau_{1.5} = {\pi\over4}(\pi Mf_0)^{-2/3}\eta^{-1},
\end{eqnarray}
\end{mathletters}
which can be inverted to obtain
\begin{mathletters}
\begin{eqnarray}
M &=& {5\over32\pi^2f_0}{\theta^2\over\theta^1},\\
\eta &=& \left[{16\pi^5\over25}{(\theta^1)^2\over(\theta^2)^5}\right]^{1/3}.
\end{eqnarray}
\end{mathletters}
This choice of $(\theta^1,\theta^2)$ lets us write the waveform phase
analytically while keeping the metric components from varying too
strongly; therefore it is convenient for calculating numbers of
templates (see below).  However, for other purposes $(\tau_0,\tau_1)$
are just as convenient, and to be consistent with the literature we
will use them.

\subsection{Noise spectra}
\label{sec:noise spectra}

In this paper we consider the noise spectra of the four large- and
intermediate-scale interferometer projects, LIGO, VIRGO, GEO600, and
TAMA.  For LIGO we use three noise spectra corresponding to three
interferometer configurations, the ``first
interferometers''~\cite{LIGO} (which are planned to perform a
gravitational-wave search in 2002--2003), the ``enhanced
interferometers''~\cite{enhanced} (which are likely to be carrying out
searches in the mid 2000's), and the ``advanced
interferometers''~\cite{LIGO} (which are thought representative of the
type of detector that might operate a few years after the enhanced
ones).  For convenience we abbreviate these, respectively, as LIGO~I,
LIGO~II, and LIGO~III.  We use the VIRGO and GEO600 noise
spectra given in Refs.~\cite{VIRGO,GEO}.  For TAMA we use the noise
spectrum given by M.-K. Fujimoto in Ref.~\cite{GRASP}.

It is desirable to have simple analytical fits to the noise power
spectral densities used.  Kip Thorne and Scott Hughes (private
communications) have provided us with fits to LIGO~I and LIGO~III,
respectively, and a fit to LIGO~II was derived in Ref.~\cite{r-modes}.
We have constructed our own fits to the remaining noise spectra listed
in the previous paragraph.  All of these analytical fits to the noise
spectra are tabulated in Table~\ref{table:noise}.

\begin{table}
\caption{
Analytical fits to noise power spectral densities $S_h(f)$ of the
interferometers treated in this paper.  Here $S_0$ is the minimum
value of $S_h(f)$, and $f_0$ is the frequency at which the minimum
value occurs.  For our purposes $S_h(f)$ can be treated as infinite
below the seismic frequency $f_s$.  The high-frequency cutoff $f_u$ is
chosen so that the loss of signal-to-noise ratio due to finite
sampling rate $2f_u$ is $0.75\%$ (see text).}
\begin{tabular}{lccrrr}
Detector & Fit to noise power spectral density & $S_0$ (Hz$^{-1}$) &
$f_0$ (Hz) & $f_s$ (Hz) & $f_u$ (Hz) \\
\hline
LIGO I & $S_0/3\,\left[(f_0/f)^4+2(f/f_0)^2\right]$ &
$8.0\times10^{-46}$ & 175 & 40 & 1300 \\
LIGO II & $S_0/11\,\left\{2(f_0/f)^{9/2}+9/2[1+(f/f_0)^2]\right\}$ &
$7.9\times10^{-48}$ & 110 & 25 & 900 \\
LIGO III & $S_0/5\,\left\{(f_0/f)^4+2[1+(f/f_0)^2]\right\}$ &
$2.3\times10^{-48}$ & 75 & 12 & 625 \\
VIRGO & $S_0/4\,\left[290(f_s/f)^5+2(f_0/f)+1+(f/f_0)^2\right]$ &
$1.1\times10^{-45}$ & 475 & 16 & 2750 \\
GEO600 & $S_0/5\,\left[4(f_0/f)^{3/2}-2+3(f/f_0)^2\right]$ &
$6.6\times10^{-45}$ & 210 & 40 & 1450 \\
TAMA & $S_0/32\,\left\{(f_0/f)^5+13(f_0/f)+9[1+(f/f_0)^2]\right\}$ &
$2.4\times10^{-44}$ & 400 & 75 & 3400
\end{tabular}
\label{table:noise}
\end{table}

The shape of each noise spectrum determines natural low- and
high-frequency cutoffs for the matched filtering integrals.  The
low-frequency cutoff $f_s$ is the defined as the frequency above which
$99\%$ of $S^2/N^2$ is obtained; we call this the {\it seismic
frequency} and denote it with a subscript $s$ because it is typically
near the frequency at which seismic noise causes the noise power
spectral density $S_h(f)$ to begin rising sharply. Because integration
below this frequency contributes little to the detectability of
signals and costs much in terms of total number of templates and
computational resources, we assume the templates to begin with
gravitational-wave frequency $f_s$.

The high-frequency cutoff $f_u$, as pointed out by \'Eanna Flanagan
(private communication), needs to be high enough that the $S/N$
degradation due to discrete time steps $\Delta t = 1/(2f_u)$ in the
data analysis is less than that due to discrete choices of the
templates' intrinsic parameters.  Quantitatively, this requires that
\begin{equation}
\gamma_{00}/(4f_u)^2 < 1-MM,
\end{equation}
where $\gamma_{00}$ is the $t_c$-$t_c$ component of the metric before
$t_c$ is projected out [see Ref.~\cite{Owen96}, Eq.~(2.23)].  At any
post-Newtonian order we have $\gamma_{00} = (2\pi f_0)^2
[J(1)-J(4)^2]/2$ [see Ref.~\cite{Owen96}, Eq.~(2.27)] and thus
\begin{equation}
f_u > \pi f_0 \sqrt{{J(1)-J(4)^2 \over 8(1-MM)}}.
\end{equation}
We have chosen $f_u$ to be twice the right-hand side of this
expression so that the loss due to sampling is $1/4$ the loss due to
discrete values of the intrinsic parameters.  There is probably a more
clever way of optimizing $f_u$, but this is a first cut.

\subsection{Number of templates and computational cost}
\label{sec:number of templates actual}

We now proceed with the calculation of the number of templates needed
to perform a single-pass, on-line search for gravitational-wave
signals of the form in Eq.~(\ref{Psi}), for the noise spectra in
Table~\ref{table:noise}.

The first task is to obtain the intrinsic-parameter metric.  We
rewrite the waveform phase at 2PN order as
\begin{eqnarray}
\label{phase}
\Psi &=& {3\over5}\theta^1\left({f\over f_0}\right)^{-5/3}
+\left[{11\pi\over12}{\theta^1\over\theta^2}
+{743\over2016}\left({25\over2\pi^2}\right)^{1/3}(\theta^1)^{1/3}
(\theta^2)^{2/3}\right]\left({f\over f_0}\right)^{-1}
-{3\over2}\theta^2\left({f\over f_0}\right)^{-2/3}\nonumber\\
&&+\left[{617\pi^2\over384}{\theta^1\over(\theta^2)^2}
+{5429\over5376}\left(25\pi\over2\right)^{1/3}\left({\theta^1\over
\theta^2}\right)^{1/3}
+{15\,293\,365\over10\,838\,016}\left({5\over4\pi^4}\right)^{1/3}
{(\theta^2)^{4/3}\over(\theta^1)^{1/3}}\right]\left({f\over f_0}
\right)^{-1/3}.
\end{eqnarray}
From Eqs.~(\ref{phase}) and Eqs.~(\ref{gamma_ab})--(\ref{project}) it
is straightforward to derive the metric components $g_{ij}$ with
symbolic manipulation software.  However, the general expressions for
the $g_{ij}$ (and even the $\gamma_{ij}$) are too lengthy and
complicated to display here.

Next, we obtain the proper volume
\begin{equation}
V=\int d^2\theta\,\sqrt{\det{\|g_{ij}\|}}
\end{equation}
of the region of interest.  The boundaries of this region are set by
the range of masses~$[m_{\min},m_{\max}]$ of the individual objects in
binaries.  It turns out that only $m_{\min}$ has a strong influence on
$V$~\cite{SD91,DS94,Owen96}.  For 1PN waveforms the intrinsic
parameter space is flat (i.e.\ $\det{\|g_{ij}\|}$ is constant) and
thus $V$ can be obtained analytically using the coordinates
$(\tau_0,\tau_1)$.  Beyond 1PN order in the waveform, the intrinsic
parameter space is not flat, and thus it is convenient to calculate
the proper volume numerically, e.g.\ by a Monte Carlo method.  In
order to make the integrand as nearly constant as possible we use the
metric tensor transformation law to switch from $(\theta^1,\theta^2)$
coordinates to $(\tau_0,\tau_1)$ [because in these coordinates the
metric components vary more slowly over the region of interest than in
$(\theta^1, \theta^2)$].  Finally, to obtain the number of templates
we divide the proper volume $V$ by the proper volume per template
$\Delta V$.  For a rectangular lattice in two dimensions, $\Delta V =
2(1-MM)$ [cf.\ Eq.~(\ref{eq:dV})].  In Table~\ref{table:calN} we give
values of ${\cal N}$ calculated for the noise spectra given in
Table~\ref{table:noise} assuming a rectangular lattice.  For a
hexagonal lattice, $\Delta V$ is given by Eq.~(\ref{eq:hex}),
resulting in a reduction of about $20\%$ in ${\cal N}$, ${\cal P}$,
and ${\cal S}$.  However, in practice much of this reduction may be
offset by the details of actually constructing a lattice (see Sec.~IV)
and therefore we make our estimates for the more conservative
rectangular case.

\begin{table}
\caption{
Numbers of templates required to cover parameter space at a minimal
match of $97\%$ with a rectangular lattice.  The region of interest is
that inhabited by binaries with component masses greater than
$m_{\min}$.  We use restricted post-Newtonian (PN) templates whose
phase evolution is accurate to the indicated order.}
\begin{tabular}{lcccc}
& \multicolumn{3}{c}{$m_{\min}=0.2M_\odot$} & $m_{\min}=1M_\odot$ \\
Detector & 1PN & 1.5PN & 2PN & 2PN \\
\hline
LIGO I &
$2.3\times10^5$ & $5.6\times10^5$ & $4.8\times10^5$ & $1.1\times10^4$ \\
LIGO II &
$1.0\times10^6$ & $1.9\times10^6$ & $1.7\times10^6$ & $4.0\times10^4$ \\
LIGO III &
$5.7\times10^6$ & $7.9\times10^6$ & $6.9\times10^6$ & $1.7\times10^5$ \\
VIRGO &
$5.8\times10^6$ & $1.1\times10^7$ & $9.3\times10^6$ & $2.2\times10^5$ \\
GEO600 &
$4.2\times10^5$ & $9.7\times10^5$ & $8.3\times10^5$ & $1.9\times10^4$ \\
TAMA &
$5.1\times10^4$ & $1.7\times10^5$ & $1.4\times10^5$ & $3.1\times10^3$
\end{tabular}
\label{table:calN}
\end{table}

These numbers are easily translated into the computational costs shown
in Tables~\ref{table:cost1} and~\ref{table:cost2}:  To a good
approximation, the length of the longest filter (with
$m_1=m_2=m_{\rm min},$ i.e. equal mass binaries so that $\eta=1/4$) 
is given by
\begin{eqnarray}
F &\simeq& 2f_u\tau_0(f_0/f_s)^{8/3}\nonumber\\
&=& {5\over32}f_u(\pi f_s)^{-8/3}(2m_{\min})^{-5/3}.
\end{eqnarray}
The storage required for all of the templates is then roughly given by
\begin{equation}
{\cal S} \simeq {\cal N} F,
\end{equation}
and the computational power ${\cal P}$ is given by Eq.~(\ref{eq:calP}).

\begin{table}
\caption{
Computational costs obtained from the numbers of templates given in
Table~\protect\ref{table:calN} using 2PN templates and $m_{\min} =
0.2M_\odot$.  The symbol $F$ denotes the length (in real numbers) of
the longest template.}
\begin{tabular}{lccc}
Detector & $\log_2 F$ & CPU power ${\cal P}$ (flops) &
Storage ${\cal S}$ (reals) \\
\hline
LIGO I & 21 & $9.9\times10^{10}$ & $1.0\times10^{12}$ \\
LIGO II & 22 & $2.5\times10^{11}$ & $7.1\times10^{12}$ \\
LIGO III & 25 & $7.8\times10^{11}$ & $2.3\times10^{14}$ \\
VIRGO & 26 & $4.8\times10^{12}$ & $6.2\times10^{14}$ \\
GEO600 & 21 & $1.9\times10^{11}$ & $1.8\times10^{12}$ \\
TAMA & 20 & $7.3\times10^{10}$ & $1.5\times10^{11}$
\end{tabular}
\label{table:cost1}
\end{table}

\begin{table}
\caption{
Computational costs as in Table~\protect\ref{table:cost1}, except
here we assume $m_{\min} = 1M_\odot$.}
\begin{tabular}{lccc}
Detector & $\log_2 F$ & CPU power ${\cal P}$ (flops) &
Storage ${\cal S}$ (reals) \\
\hline
LIGO I & 17 & $1.9\times10^9$ & $1.4\times10^9$ \\
LIGO II & 19 & $5.2\times10^9$ & $2.1\times10^{10}$ \\
LIGO III & 21 & $1.7\times10^{10}$ & $3.6\times10^{11}$ \\
VIRGO & 22 & $9.8\times10^{10}$ & $9.1\times10^{11}$ \\
GEO600 & 17 & $3.7\times10^9$ & $2.5\times10^9$ \\
TAMA & 16 & $1.3\times10^9$ & $2.0\times10^8$
\end{tabular}
\label{table:cost2}
\end{table}

\subsection{Scaling laws}

Approximate scaling laws can be obtained as follows.  At 1PN order the
intrinsic parameter space is flat, and thus the proper volume of the
region of interest scales as the (dimensionless) coordinate volume, i.e.
\begin{equation}
V \sim f_0\tau_0 \times f_0\tau_1 \sim m_{\min}^{-8/3} f_0^{-8/3}.
\end{equation}
At higher post-Newtonian orders the metric components with respect to
the $(\tau_0,\tau_1)$ coordinates are nearly constant, and thus this
scaling still roughly holds.

From the scaling of the proper volume we can obtain the scalings of
the other quantities of interest.  Inserting the
dependence~(\ref{eq:dV}) of the proper volume per template on the
minimal match, we obtain
\begin{equation}
{\cal N} \sim (1-MM)^{-1} m_{\min}^{-8/3} f_0^{-8/3}.
\end{equation}
Taking Eq.~(\ref{eq:calP}), neglecting the weak logarithmic
dependence, and noting that $f_u$ is proportional to $f_0$, we find
that
\begin{equation}
{\cal P} \sim (1-MM)^{-1} m_{\min}^{-8/3} f_0^{-5/3}.
\end{equation}
Multiplying the number of templates ${\cal N}$ by the length of each
template $F$, the storage requirement scales as
\begin{equation}
{\cal S} \sim (1-MM)^{-1} m_{\min}^{-13/3} f_0^{-5/3} f_s^{-8/3}.
\end{equation}
This last scaling with $m_{\min}$ presents a significant obstacle to
efficiently searching for binaries composed of low-mass objects, such
as MACHOs if they are low-mass ($\sim.5M_\odot$) black
holes~\cite{machos}.  Searches for low-mass objects will likely need
to generate templates as needed rather than store them, incurring
additional CPU costs not addressed in this paper.

Note that the above scaling laws implicitly assume that $f_0$ or $f_s$
is varied while holding the overall shape of the noise spectrum fixed.
The noise spectrum of a real interferometer, being composed of many
independent noise sources, is unlikely to change in such a manner
except for fairly small changes in $f_0$ or $f_s$.  However, these
scaling laws give one a rough feel for (i) how changes in a single
interferometer can affect data analysis requirements, and (ii) why
different interferometers have drastically different requirements.

\section{A Template Placement Algorithm}
\label{sec:placement}

\begin{figure}
\centerline{\psfig{file=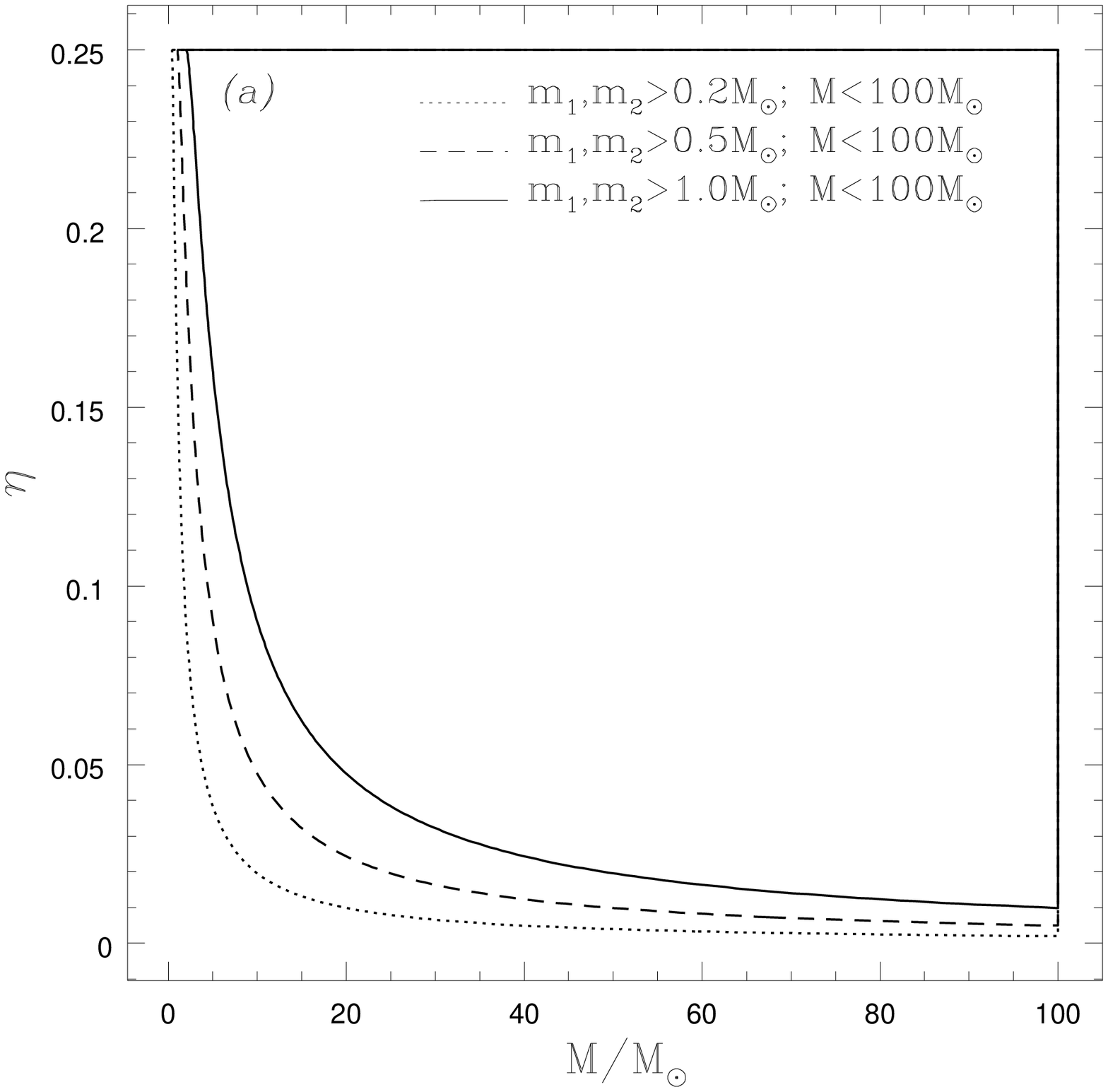,height=3.in}}
\centerline{\psfig{file=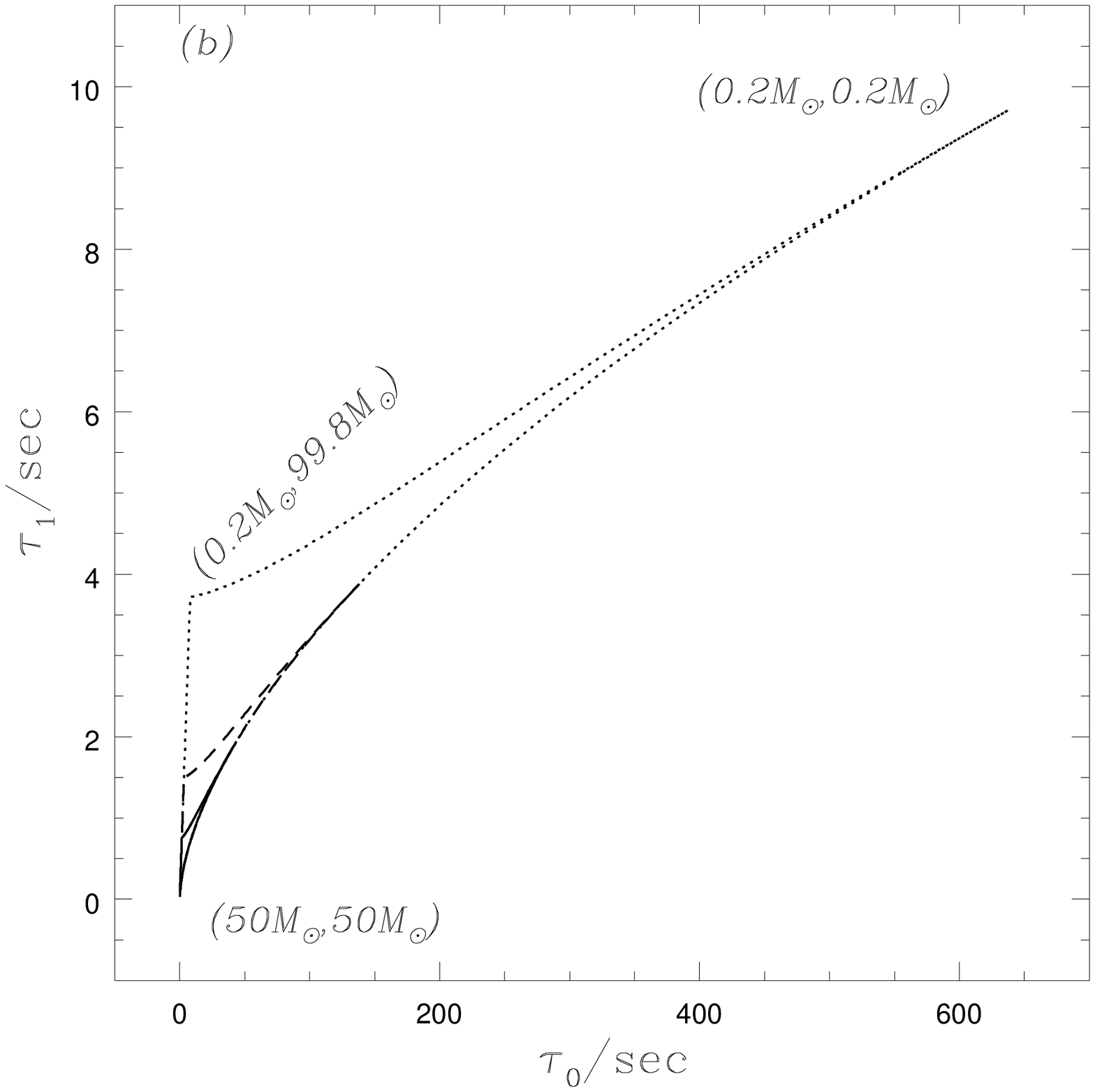,height=3.in}}
\caption{
The parameter space of search in terms of (a) standard binary mass
parameters and (b) chirp times, for different values of the lower mass
limits.}
\label{fig:parameter_space}
\end{figure}

In this Section we deal with the actual placement of the templates in
the parameter space. The parameter space is shown in
Fig.~\ref{fig:parameter_space}a in terms of the total mass $M$ and
symmetric mass ratio $\eta$ and in Fig.~\ref{fig:parameter_space}b in
terms of chirp times $\tau_0$ and $\tau_1,$ for searches in which the
maximum total mass $M_{\max}$ of the system is 100 $M_\odot$ and the
lower limit $m_{\min}$ on the mass of each component star is $0.2,$
$0.5$ or $1.0 M_\odot.$ The bottom line in
Fig.~\ref{fig:parameter_space}a corresponds to binaries of equal mass
$(\eta=1/4)$ with the rightmost point corresponding to lowest mass
binaries and the leftmost to greatest mass binaries of our search.
There are no binaries in the region below this line as the parameter
$\eta$ exceeds 1/4 there, which is unphysical.  The top and the left
lines are determined by $m_{\min}$ and $M_{\rm max}$ respectively.
Given the minimum mass of the component stars and the maximum total
mass, the parameter space of binaries is completely fixed.  The volume
of the parameter space (and the corresponding number of templates
required) is a sharp function of the lower cutoff in the masses of the
component stars and increases, as we have seen, as $m_{\rm
min}^{-8/3}.$

\begin{figure}
\centerline{\psfig{file=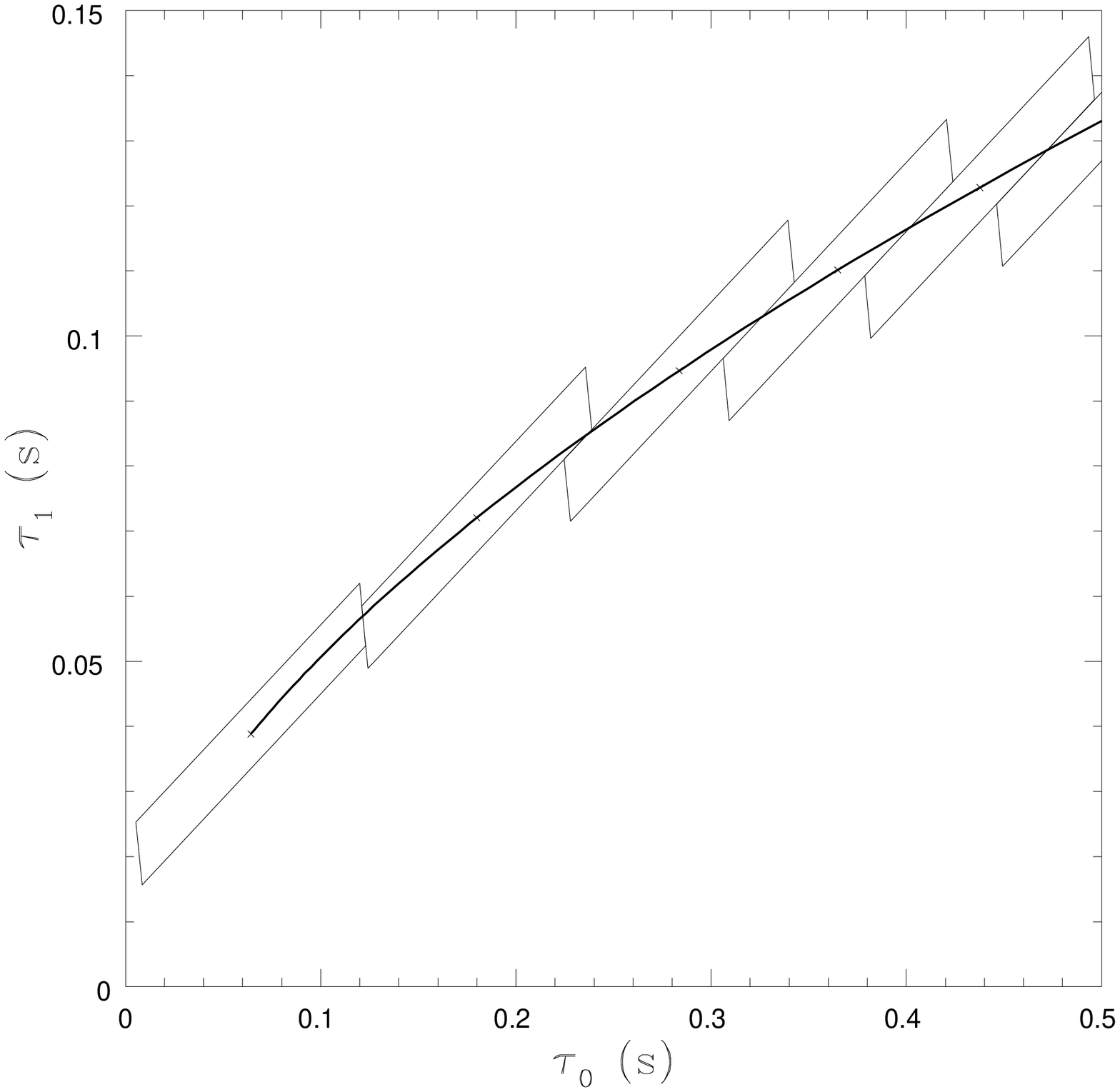,height=3.in}}
\caption{  
Illustration of optimal translation of a template along an arbitrary
curve.}
\label{fig:translation}
\end{figure}

The shape of the parameter space is rather complicated and attention
needs to be paid in the placement of templates so that the inevitable
{\it spill over} (see below) is minimal. Our implementation of the
filter placement is motivated by the following astrophysical
consideration: The observed neutron stars are all of roughly equal
mass~\cite{taylor}.  It is therefore to be expected that many inspiral
signals will come from equal mass binaries.  Consequently, we optimize
the filter placement for equal mass binaries. This is achieved by
beginning our template placement along the $\eta=1/4$ line. The span
of each template is taken to be the largest rectangle (in a coordinate
system in which the metric is locally diagonal) that can be inscribed
inside the minimal match ellipse (we take $MM=97\%$).  Note that in
Fig.~\ref{fig:translation} the spans do not appear rectangular because
they are sheared by transforming from coordinates in which the metric
is locally diagonal (see below) to the $(\tau_0,\tau_1)$ coordinates.
We begin with the leftmost point on the bottom edge of the parameter
space of Fig.~\ref{fig:parameter_space}b, corresponding to the most
massive binary of our search with the shortest chirp time. Such a
system will, of course, consist of equal mass bodies and therefore our
starting point is on the $\eta=1/4$ curve. The next template is placed
on the $\eta=1/4$ curve at that location where the left edge of its
rectangle touches the right edge of the previous template's rectangle.
In a sense this is a straightforward generalization of placement of
filters discussed in Ref.~\cite{Owen96} along grid lines that are not
necessarily straight.

This optimal translation of templates is most easily done in a
coordinate system in which the metric is locally diagonal.  Let
$(x_0,x_1)$ denote such a coordinate system (found by diagonalizing
the two-dimensional matrix $g_{ij}$) and let $f(x_0,x_1)=0$ denote a
curve in the two-dimensional $x_0$--$x_1$ plane along which templates
are to be placed.  For instance, $\eta=1/4$ in
Fig.~\ref{fig:parameter_space} is one such curve.  A convenient point
to begin is the point $(x_0^{(1)}, x_1^{(1)})$ at one end of the
curve.  In Fig.~\ref{fig:translation} we have sketched an arbitrary
curve together with the first template and its span.  The span of a
template, for minimal matches close to 1, is an ellipse.  However, its
effective (non-overlapping) span, when setting up a lattice of
templates, is only an inscribed polygon such as a rectangle or an
irregular hexagon.  In the following discussion for simplicity we
consider the span to be a rectangle and hence we will be setting up a
rectangular lattice.  By choosing a hexagonal lattice the number of
templates can be reduced by about $20\%$, but the reduction is less
when the curve along which templates need to be placed is parallel to
neither $x_0$ nor $x_1$ axis.

Given the `local' distance $(dx_0^{(1)}, dx_1^{(1)})$ between
templates, we can get two points on the curve $f(x_0, x_1) = 0$ which
are simultaneous solutions of
\begin{equation}
\{f(x_0,x_1)=0, \quad x_n=x_n^{(1)}+dx_n^{(1)} \}
\end{equation}
for $n=0$ and $n=1$ respectively.  In order to cover the parameter
space without leaving any `holes' it is obvious that the next template
should be placed at the point that is nearer to the first
template. This is how one obtains the nearest neighbour of a template.

\begin{figure}
\centerline{\psfig{file=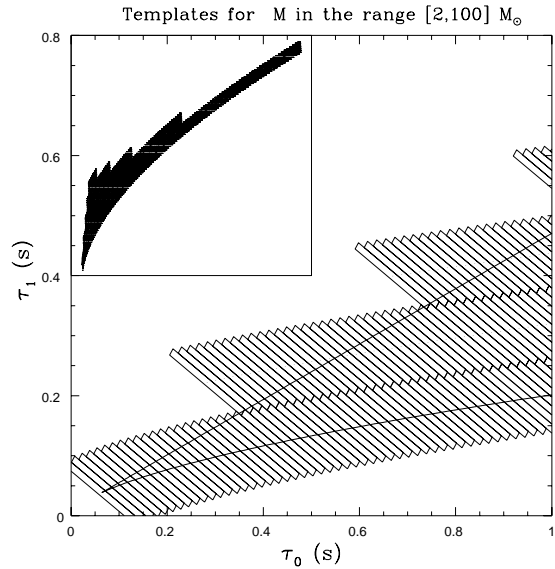,height=3.in}}
\caption{
Choice of templates in the space of chirp times for initial LIGO
interferometer with an upper frequency cutoff of 2 kHz and for a
search of binaries with masses of component stars larger than $1
M_\odot$ and total mass no greater than $100 M_\odot$.  The parameters
characterizing the lattice at the location $(M=1.4 M_\odot,\eta=1/4)$
are $dx_0=46.5$~ms, $dx_1=4.45$~ms and $\theta=0.35$~rad, and vary
slightly from these values at other locations.}
\label{fig:placement}
\end{figure}

Returning to our problem of placing templates in the parameter space
of chirp times, we move along the $\eta=1/4$ line till the binary of
longest chirp time is reached. The set of templates chosen on the
$\eta=1/4$ curve form the base, on top of which we construct layers to
fill the parameter space.  The next row of templates is set on top of
the first so that the region of interest is completely covered by the
spans---which means that the first few templates will be located
outside that region, and that the total number of templates must be
modestly greater than naively estimated by Eq.~(\ref{eq:calN}) (see
Fig.~\ref{fig:placement}).

\section{Discussion}
\label {sec:discussion}

In this paper we have discussed the problem of optimally placing
templates in a binary inspiral signal search.  For our templates we
have used the restricted second post-Newtonian waveform.  We have made
estimates of the computational costs of an on-line search of the
inspiral waveform for all the large- and mid-scale interferometeric
detectors now under construction.  We have addressed several important
issues: (i) the density of templates in the parameter space, (ii) the
set of parameters most suitable for easy placement of templates, and
(iii) the number of templates and computational resources needed to
analyze the data on-line.

These estimates should serve as a baseline for explorations of other
data analysis strategies, and some of the techniques here could be
incorporated into other strategies.  For example, it is now recognized
that a substantial reduction in computational cost can be achieved by
carrying out a hiearchical search~\cite{MD96,Mohanty98}.  For a
two-step hierarchical search strategy, in the first step a sparsely
filled family of templates is used, with a threshold lower than what
is acceptable based on the expected number of false alarms.  Those
events which cross this trigger threshold are further examined with a
finer grid of templates chosen around the template that triggered the
event.  In such a hierarchical search, templates chosen in the first
step will essentially be the same for each data segment.  However,
templates in the second step need to be changed from one data segment
to the next, depending on which templates from the coarse grid family
produces an `event'.  It is in the case of a coarse grid that our
analytical algorithm for template placement and analytical estimates
of computational requirements fail and must be replaced by numerical
methods that are computationally expensive.  But generating filters
corresponding to the first step of the hierarchical search is more or
less a one-time job.  A finer grid is to be chosen quite frequently
(essentially each time a possible event is selected in the first step)
and in this case, fortunately, the analytical techniques discussed in
this paper are quite accurate and one does not have to follow the time
consuming numerical placement of templates either for estimating
computational costs or for actually performing the search.

There are several important problems we have not addressed in this
paper which could be the topics of future work.  We have not treated
the problem of searching the corners of parameter space where the
precessing binaries live.  Although requiring some effort to seek out,
these systems could prove quite informative and astrophysically
interesting.  The problem of searching for precessing binaries has
been addressed only in a very exploratory way~\cite{A96} but could
benefit from further analysis using the techniques of this paper.
Also, we have used a crude relation between the minimal match and
fraction of event rate lost.  This could be improved by a statistical
analysis such as begun by Mohanty~\cite{Mohanty98}.  Now that the
``P-approximants''~\cite{DIS98} have proven a promising way of
building templates, it is important to examine the computational costs
of using them to conduct a search.

\acknowledgments

Both authors were partially supported by NSF Grant PHY-9424337.  BJO
was also supported by the NSF graduate program.  One of us (BSS) would
like to thank Bernard Schutz for hospitality at the Albert Einstein
Institute in the summer of 1997 where some of this work was carried
out. We are indebted to
Bruce Allen, Kent Blackburn, Tom Prince, and Kip Thorne for fruitful
conversations, and we thank many members of the LIGO, VIRGO, GEO600,
and TAMA projects for helping us with their respective noise curves.


\begin{references}
\bibitem{Wiener}
Some physicists are in the habit of calling this optimal filtering
method simply ``Wiener filtering.''  However, in the larger signal
processing community, ``Wiener filtering'' refers to an altogether
different technique.  In order to avoid confusion we call this method
``matched filtering.''
\end{references}
\end{document}